\documentclass[pre,amsmath,amssymb,floatfix,preprint,showpacs]{revtex4}
\usepackage{graphicx}
\usepackage{float}
\usepackage{bm}
\usepackage[utf8]{inputenc}

\begin{document}
\title{The condensation and ordering of models of empty liquids}
\author{N. G. Almarza}
\affiliation{Instituto de Qu{\'\i}mica-F{\'\i}sica Rocasolano, CSIC, Serrano 119, E-28006 Madrid, Spain }
\author{J. M. Tavares}
\affiliation{Centro de F\'{\i}sica Te\'orica e Computacional, Universidade de Lisboa, Avenida Professor Gama Pinto 2,
P-1649-003 Lisbon, Portugal}
\affiliation{Instituto Superior de Engenharia de Lisboa, Rua Conselheiro Em\'{\i}dio Navarro 1, 
P-1950-062 Lisbon, Portugal}
\author{M. Sim\~oes, M. M. Telo da Gama}
\affiliation{Centro de F\'{\i}sica Te\'orica e Computacional, Universidade de Lisboa, Avenida Professor Gama Pinto 2,
P-1649-003 Lisbon, Portugal}
\affiliation{Departamento de F\'{\i}sica, Faculdade de Ci\^encias, Universidade de Lisboa, Campo Grande,
P-1749-016 Lisbon, Portugal}
\date{\today}
\begin{abstract}

We consider a simple model consisting of particles with four bonding sites (``patches''), two of type $A$ and two of type $B$, on the square lattice, and investigate its global phase behavior by simulations and theory. 
We set the interaction between $B$ patches to zero and calculate the phase diagram as the ratio between the $AB$ and the $AA$ interactions, $\epsilon_{AB}^*$, varies. In line with previous work, on three-dimensional off-lattice models, we show that the liquid-vapor phase diagram exhibits a re-entrant or ``pinched'' shape for the same range of $\epsilon_{AB}^*$, suggesting that the ratio of the energy scales - and the corresponding empty fluid regime - is independent 
of the dimensionality of the system and of the lattice structure. In addition, the model exhibits an order-disorder transition that is ferromagnetic in the re-entrant regime.
The use of low-dimensional lattice models allows the simulation of sufficiently large systems to establish the nature of the liquid-vapor critical points 
and to describe the structure of the liquid phase in the empty fluid regime, where the size of the ``voids'' increases as the temperature decreases. 
We have found that the liquid-vapor critical point is in the 2D Ising universality class, with a scaling region that decreases rapidly as the temperature decreases.   

The results of simulations and theoretical analysis suggest that the line of order-disorder transitions intersects the condensation line at a 
multicritical point at zero temperature and density, for patchy particle models with a re-entrant, empty fluid, regime. 

\end{abstract}
\maketitle

\section{Introduction}

One of the scientific and technological revolutions currently in progress is the increasing ability to miniaturize material design 
and manufacturing components. Advances in the chemical synthesis and fabrication of nanometer-to-micrometer sized particles have 
produced a variety of new particles. Their organization into more complex structures remains, however, a great challenge. A promising 
approach inspired by Nature is nanoparticle self-assembly. The structure of the self-assembled clusters, which range from chains to 
rings and complex branched structures, depends crucially on the anisotropy of the particle shapes and interactions and may compete 
with the clustering that drives condensation, giving rise to novel macroscopic behavior \cite{[11],[12],[13]}. 

Indeed, patchy particle models with dissimilar patches ($A$ and $B$) were recently introduced in this context and revealed that the criticality 
depends on the type of clusters that are formed \cite{[14],[15]}. Of particular interest are systems where the self-assembled clusters are long 
linear chains connected by junctions, as the liquid-vapor transition of these network (percolated) fluids may be viewed as the condensation of 
these junctions \cite{[14],[15]}. In addition to ferrofluids or electro-rheological fluids, colloids with distinct patchy interactions may be 
synthesized by the selective functionalisation of specific areas of the particles \cite{[16],[17]}.  

Primitive models of patchy particles with identical \cite{[18],[19]} and distinct patches share the physics of limited valence materials, 
namely the existence of stable liquid states of vanishingly small density (empty liquids), and provide a route to equilibrium gels \cite{[20]}. 
In addition, distinct-patch models allow a unique control of the effective valence through the temperature $T$. In three-dimensional (3D) off-lattice 
models consisting of particles with two types of patches, $A$ and $B$, where the interaction between $B$ patches is set to zero, the topology of the 
liquid-vapor diagram is determined by the ratio between the $AB$ and the $AA$ interactions, $\epsilon_{AB}^*$. As $\epsilon_{AB}^*$ 
decreases in the range $\frac{1}{3}< \epsilon_{AB}^* < \frac{1}{2}$, the low-temperature liquid-vapor coexistence region also decreases \cite{[21]}. 
The binodal exhibits a characteristic re-entrant or ``pinched'' shape with the coexisting liquid density vanishing as the temperature approaches zero 
\cite{[21],[22]}. Below $\epsilon_{AB}^* = \frac{1}{3}$ there is no condensation, and above $\epsilon_{AB}^* = \frac{1}{2}$ there is no 
re-entrant behavior \cite{[23]}. Both the scaling of the vanishing critical parameters and the re-entrant phase behavior are predicted correctly 
by Wertheim's thermodynamic first-order perturbation theory \cite{[21],[22],[24],[25]}. The theory also reveals that the re-entrant phase behavior 
is driven by the balance of two entropic contributions: the higher entropy of the junctions and the lower entropy of the chains in the (network) liquid 
phase, as suggested a decade ago on the basis of a hierarchical theory of network fluids \cite{[26]}.

The feature that makes patchy particles ideally suited to the investigation of the interplay between self-assembly and condensation is the fact 
that both the thermodynamic and structural properties of patchy particle systems can be predicted with a high degree of accuracy by the thermodynamic perturbation theory of Wertheim and the Flory-Stockmayer theory of polymerization \cite{[27],[28],[29]}. It is then possible to study the phase behavior 
of patchy particles using reliable liquid-state theories and to use this knowledge to design the models and guide the simulations, the results of which 
validated the theoretical predictions \cite{[18],[19],[21],[22]}. There remain, however, two open questions: 1. What is the nature of the liquid-vapor 
critical point, in models with an empty fluid regime ? and 2. Are there ordered phases that pre-empt the empty fluid regime or, What is the topology 
of the global phase diagram ? These are difficult questions that will be addressed here by considering simple patchy particle models on the square lattice.  

In systems with two bonding sites per particle, only (polydisperse) linear chains form and there is no liquid-vapor phase transition \cite{[30]}. 
If the chains are sufficiently stiff they undergo an ordering transition at fixed concentration, as the temperature decreases below the bonding temperature. 
The interplay between the self-assembly process, driven by the bonding interactions, and the ordering transition, driven by the anisotropic 
shape of the bonded clusters has been investigated for a two-dimensional (2D) model consisting of particles with two bonding sites, on the square lattice (self-assembling rigid rods or SARR model). It was shown that bonding drives ordering and that the ordering enhances bonding \cite{[31]}. Subsequently, 
extensive Monte Carlo simulations were carried out to investigate the nature of the ordering transition that was shown to be in the Ising 2D universality 
class, as in models where the rods are monodisperse \cite{[32]}. The scaling region, however, was found to depend strongly on the temperature \cite{[32],[33]}. 

In this paper, we consider the $2A2B$ model consisting of particles with four patches, two of type $A$ and two of type $B$, on the square lattice and 
investigate its global phase behavior by simulations and theory. We set the interaction between $B$ patches to zero and calculate the phase diagram, 
as the ratio of the $AB$ and the $AA$ interactions, $\epsilon_{AB}^*$, varies. We find that, in the same range of parameters as in 3D off-lattice models, 
the liquid-vapor diagram exhibits a re-entrant or ``pinched'' shape, and there is an empty fluid regime. 
In addition, below $\epsilon_{AB}^* = \frac{1}{3}$ condensation ceases to exist, and the re-entrant regime disappears for $\epsilon_{AB}^*> \frac{1}{2}$, 
in line with the results for off-lattice 3D models and the predictions of Wertheim's theory \cite{[21],[22],[23]}. This suggests that the thresholds 
predicted by Wertheim's theory are exact and universal, i.e. independent of the dimensionality of the system and of the lattice structure. 
Finally, the $2A2B$ model exhibits an order-disorder (O-D) transition that is ferromagnetic for $ \frac{1}{3} <\epsilon_{AB}^*< \frac{1}{2} $.

The use of 2D lattice models allows the simulation of larger systems enabling us to establish the nature of the critical points and to investigate 
the structure of the network liquid phase in the empty fluid regime, where the size of the ``voids'' increases rapidly as the temperature decreases. 
We find that the liquid-vapor critical points are in the 2D Ising universality class, with a scaling region that decreases as the temperature decreases. 
The simulation results also indicate that the line of O-D transitions intersects the condensation line at zero temperature and density, at a multicritical 
point, or at a very low temperature, at a critical end-point. The analysis of this region requires the simulation of larger systems at extremely low temperatures, which becomes prohibitive even for 2D patchy particle lattice models. 

In order to proceed we consider a low-temperature-model (LTM) that describes the asymptotic behavior of 2D patchy particle models at low temperatures, 
and use a cluster algorithm that enables the efficient simulation of these low temperature systems. Finally, we derive asymptotic expressions based 
on Wertheim's theory, for the liquid branch of the binodal and the O-D transition that suggest, in line with the simulation results, that the transition 
lines meet, at a multicritical point, at zero temperature and density.    

The paper is arranged as follows: In section II we describe the patchy particle model, the mapping of the full lattice limit and the simulation methods. 
In section III we present the results for the global phase diagram of a system with a re-entrant binodal. We compute the binodals, analyze the nature 
of the liquid-vapor critical points, and discuss the topology of the global phase diagram for systems with a ferromagnetic ordering transition (re-entrant regime). In section IV we introduce the LTM and the simulation techniques developed to sample low temperatures efficiently. We compute the binodals and the ferromagnetic ordering transition and discuss the topology of the global phase diagram. Then, in section V we address the zero temperature and zero 
density limit theoretically. We derive asymptotic expressions for the condensation and O-D transitions based on Wertheim's theory for associating liquids, 
in the limit of strong $AA$ bonding. We conclude, based on the asymptotic analysis, that the condensation and O-D lines meet at a multicritical point, at 
zero temperature and density. In section VI we make some concluding remarks and in the Appendix provide details of the calculation of the starting point 
of the liquid-vapor equilibrium of the LTM, used in the Gibbs-Duhem integration of the liquid branch of the binodals.   

\section{The $2A2B$ model}

\begin{figure}
\includegraphics[width=100mm,clip=]{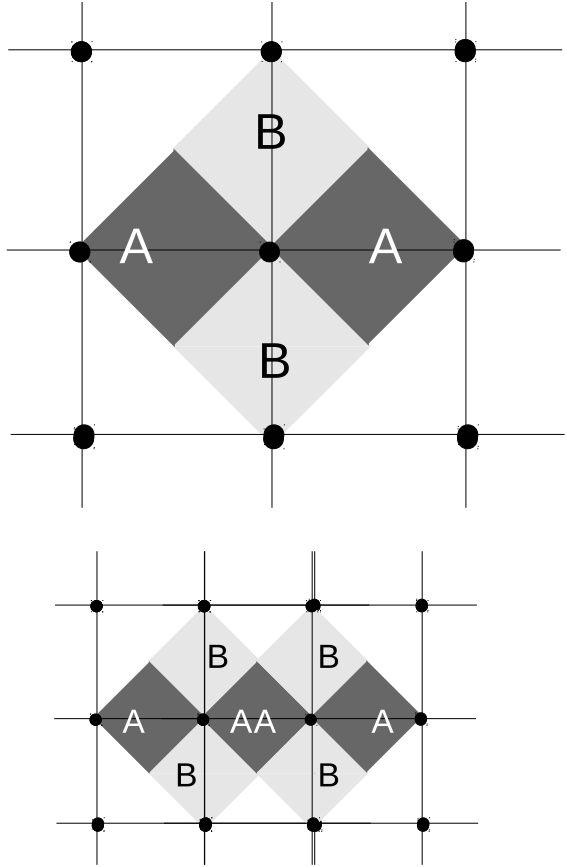}
\caption{Illustration of the model. Top: One particle with 2$A$ and 2$B$ bonding sites or patches
with the $A$ patches aligned along $\hat x$ and the $B$ patches aligned along $\hat y$.
Bottom: Two particles forming a $AA$ bond along $\hat x$.}
\label{Fig01}
\end{figure}

The model consists of particles with four patches, two of type $A$ and two of type $B$, on a square lattice. The lattice sites 
are either empty or occupied by one single particle. The patches $A$ and $B$ are aligned along one of the two lattice directions 
(See Figure \ref{Fig01}). There are two configurations for each occupied site:
(1) $A$ patches aligned along $\pm \hat{x}$ and $B$ patches aligned along $\pm \hat{y}$ , and the symmetric configuration with 
(2) $A$ patches aligned along $\pm \hat{y}$ and $B$ patches aligned along $\pm \hat{x}$.
The potential energy, ${\cal U}$, is the sum of pair interactions between nearest-neighbor (NN) particles on the lattice and is
written as:
\begin{equation}
{\cal U} = - \epsilon_{AA} {\cal N}_{AA} - \epsilon_{AB} {\cal N}_{AB} - \epsilon_{BB} {\cal N}_{BB};
\label{eq1}
\end{equation}
where ${\cal N}_{\alpha\beta}$ is the number of $\alpha\beta$ bonds, i.e. lattice bonds between NN occupied sites connecting patches 
$\alpha$ and $\beta$.   

This model is a lattice realization of the patchy particle models with distinct patches introduced in \cite{[14],[15]} and 
investigated in the context of empty network fluids \cite{[21],[22]}. In line with previous work we take $\epsilon_{AA}=\epsilon$ 
as the energy scale ($\epsilon > 0$), and focus on systems where the $B$ patches do not interact, i.e. $\epsilon_{BB} = 0$. The interaction 
between $A$ and $B$ patches varies although most of the results are for systems in the re-entrant regime, i.e. $0<\epsilon_{AB}\le\epsilon/2$. Taking into account that each particle carries two $A$ patches and that a patch can participate, at most, in one bond we can write:
\begin{equation}
2 N = 2 {\cal N}_{AA} + {\cal N}_{AB} + {\cal N}_{A0};
\label{eq2}
\end{equation}
where $N$ is the number of particles in the system and ${\cal N}_{A0}$ is the number of patches that are not bonded.  Combining Eqs. (\ref{eq1}) and (\ref{eq2}) for $\epsilon_{BB}= 0$ we get:
\begin{equation}
{\cal U}/\epsilon  =  - N + {\cal N}_{AB} \left( \frac{1}{2} - \frac{\epsilon_{AB}}{\epsilon} \right)
+ \frac{1}{2}  {\cal N}_{A0}.
\label{eq3}
\end{equation}
 From Eq. (\ref{eq3})  It follows that $AA$ bonds are favored when $\epsilon_{AB}<\epsilon/2$, while $AB$ bonds are favored 
when $\epsilon_{AB} > \epsilon/2$. At low temperature,  most of the $A$ patches are bonded and the network fluid consists of $AA$ chains connected by a small number of $AB$ branches 
in the former case while the network is almost fully branched in the latter. The special case $\epsilon_{AB}= 0$ corresponds to a self-assembling 
rigid rod (SARR) model that was studied on 2D lattices recently \cite{[31],[32],[33],[34],[35],[36]}. In the SARR model, a continuous O-D 
transition is found to be the only feature of the phase diagram. At low temperatures the particles form long rigid rods, through $AA$ bonds, 
which undergo an orientational ordering transition, in the 2D Ising class on the square lattice and in the q=3 Potts class on the triangular one 
\cite{[32],[33]}. The SARR model has no liquid-vapor transition as adjacent rods do not interact energetically.

\subsection{The full lattice limit}

The full lattice limit of the SARR model on the square lattice has been mapped on to the Ising model \cite{[32],[36]}. This was achieved by 
establishing a correspondence between the particle orientations of the SARR model and the spins $\pm 1$ of the Ising model. The total energy of 
both models is then computed by adding the contributions of elementary plaquettes, consisting of a square with four sites enclosing an elementary 
lattice cell. The mapping between the Ising and the full lattice limit of the SARR model, is established for any plaquette configuration and the 
critical temperature of the model is identified with the exact result of the critical temperature of the corresponding Ising model \cite{[32]}.
Following this procedure a similar mapping is established for the $2A2B$ model. At full lattice occupancy, the $2A2B$ model undergoes an Ising O-D 
transition, at the reduced temperature \cite{[32]}:
\begin{equation}
\frac{k_B T_c}{\epsilon} = T_c^* = \left| \frac{ \epsilon+\epsilon_{BB} -2 \epsilon_{AB}}
{2 \epsilon \ln \left( 1 + \sqrt{2} \right) } \right|;
\label{Tc2A2B}
\end{equation}
where $k_B$ is Boltzmann's constant.
The ordered phase is stable at $T<T_c$. When $\epsilon_{BB}=0$, the ordered phase is ferromagnetic (particles aligned in the same direction) 
if $\epsilon_{AB} < \epsilon/2$, and antiferromagnetic otherwise, $\epsilon_{AB} > \epsilon/2$.

Note that when $\epsilon_{BB}=0$ and $\epsilon_{AB}=\epsilon/2$ there is no O-D transition. Inspection of Eq. (\ref{eq3}) reveals that
at full lattice occupancy every $A$ patch is bonded, and therefore all the configurations have the same potential energy. This degeneracy does not
hold when vacancies (empty sites) are present but the free energy is still dominated by entropic terms that prevent the system from ordering. 

\subsection{Simulation methods}

We aim at computing the global phase diagram of the $2A2B$ model through Monte Carlo Simulation. Based on previous results \cite{[32],[33],[21],[22]} 
a low temperature critical line corresponding to the O-D transition, the locus of which at $\rho=1$ is known exactly through the mapping to the Ising 
model (\ref{Tc2A2B}), is expected to occur; in addition, a liquid-vapor first-order transition ending at a critical point (for $\epsilon_{AB}$ above a 
certain threshold) is also expected. 

The O-D transition is located using techniques analogous to those described in \cite{[33]}. We fix the temperature and system size 
and, by means of the simulated tempering algorithm, compute the properties for different values of the chemical potential around the expected 
critical point. By using appropriate finite-size scaling analysis we obtain estimates of the critical parameters, $\mu_c(T)$ and $\rho_c(T)$, in 
the thermodynamic limit.

The liquid vapor equilibria (LVE) is computed using a combination of Wang-Landau multicanonical simulation (WLMC) \cite{Lomba_2005} and 
Gibbs-Duhem integration (GDI) procedures \cite{Kofke} adapted to lattice models \cite{Hoye,Almarza-Noya}. The WLMC methodology was described previously, 
including the details specific to lattice models \cite{Hoye,Almarza-Capitan,Almarza-Lomba-2008}. WLMC simulations, combined with finite-size analysis 
techniques, are very efficient in locating the liquid-vapor critical point, and in computing the phase diagram at temperatures not far from it. At low temperatures we found it useful to resort to GDI schemes. 

We run WLMC simulations and locate the LVE at a given temperature and system size by searching for the value of the chemical potential, 
$\mu_0(L,T)$, that maximizes the density fluctuations: $\delta \rho = \left[ < \rho^2 > - < \rho >^2 \right]^{1/2} $. 
Under these conditions we compute the average density, $\rho_m = \rho_m (L,T,\mu_0)$, and the moments of the density distribution, $m_k = < (\rho - \rho_m)^k >$, in order to calculate the ratio $g_4 = m_4 /m_2^2$, which is related to the fourth-order Binder cumulant \cite{Landau_Binder}.
We establish whether at the chosen temperature $T$ there is LVE by analyzing the dependence of $g_4$ on the system size. LVE occurs below the critical 
temperature, where at $\mu_0$ the density distribution function exhibits two peaks that become sharper as the system size $L$ increases. This implies that $g_4(L)$ decreases as $L$ increases and approaches $g_4 = 1$ in the thermodynamic limit. Above $T_c$, $g_4(L)$ increases with $L$ and approaches $g_4 = 3$ 
(Gaussian distribution) in the thermodynamic limit. At the critical temperature, finite-size scaling arguments \cite{Wilding,Perez-Pellitero,Lomba_2005}, indicate that (for sufficiently large systems) $g_4(L)$ takes a non-trivial value that depends on the boundary conditions and on the universality class of the transition. 

We estimate system size dependent pseudo-critical points: $[T_c(L),\mu_c(L)]$ by imposing that $g_4(L,T)$ takes the value corresponding to the 2D Ising universality class \cite{Salas}. Numerical details of these calculations may be found elsewhere \cite{Lomba_2005,Almarza-Lomba-2008}. We proceed to estimate the critical temperature and density in thermodynamic limit, using the scaling equations:\cite{Wilding}
\begin{equation}
\rho_c(L) - \rho_c \propto L^{-2+\frac{1}{\nu} },
\label{eq.rhoc}
\end{equation}
\begin{equation}
T_c(L) - T_c \propto L^{-\frac{1}{\nu} -\lambda } 
\label{eq.tc}, 
\end{equation}
where $\nu$ is the correlation length critical exponent ($\nu=1$ for the 2D Ising class); 
and  $\lambda = \theta/\nu$, where $\theta$ is the correction to scaling critical exponent. There is some controversy \cite{Barma,Blote,Bruce,Nienhuis,Kamienarz} concerning the value of $\lambda$ for systems in the 2D Ising class, as a number of simple models (e.g. 2D Ising) have no irrelevant operators \cite{Barma,Salas}. 
One then expects, $\lambda=4/3$ \cite{Nienhuis,Bruce} in general or $\lambda=7/4$ \cite{Kamienarz,Salas} in the absence of irrelevant operators. Taking this into account, we computed three  estimates of the critical temperature (the same scheme applies to the critical chemical potential), using: $\lambda=4/3$; $ \lambda=7/4$; and considering $\lambda$ as a fitting parameter.

At temperatures below $T_c$ we fit the system-size dependent LVE results to the scaling equations:
\begin{equation}
x(L,T) - x(T) \propto  L^{-d};
\end{equation}
where $x(L,T)$ is the finite-size result for the property $x$, and $x(T)$ (obtained from the fit) is the estimate of the property in the 
thermodynamic limit; $d=2$ is the spatial dimensionality of the system. We have obtained very precise values of the chemical potential at coexistence, 
which were subsequently used in the GDI to compute the LVE in a wider range of temperatures (away from the critical point). 
Within the GDI we run sequences of two phase (liquid and vapor) simulations using larger system sizes (than those feasible with WLMC) allowing us to sample 
lower temperatures, in the empty liquid regime.

In the computation of the critical parameters, described above, we assumed that the critical point of the LVE is in the 2D Ising universality class. 
As this is not yet established, we proceed to analyze the scaling behavior of the pseudocritical parameters, and the moments of the density distribution, 
$P_L(\rho)$, at the pseudocritical points. 
The finite-size scaling behavior of  $\delta \rho_c(L)$ satisfies:\cite{Blote_1995}
\begin{equation}
L^{\beta'/\nu}\delta \rho_c(L) \approx a_0 + a_1 L^{-\lambda};
\label{eq.ddc}
\end{equation}
where $\beta'$ is the critical exponent of the order parameter ($\beta'=1/8$ for 2D Ising).
We expect the shape of the critical density distribution, $P_L(\rho)$, to approach that of the critical Ising 2D magnetization $P^{{\text Ising}}({\cal M})$, for large system sizes \cite{Wilding}. Deviations occur for small systems due to corrections to scaling associated to irrelevant fields and field-mixing contributions\cite{Wilding}. Thus, in addition to checking the scaling of $T_c(L)$, $\rho_c(L)$ and $\delta \rho_c(L)$, we compare the asymptotic values of the reduced moments of the density distribution, $g_5=m_5/m_2^{5/2}$ and $g_6=m_6/m_2^3$, to the critical Ising 2D values, $g_5=0$ and $g_6\simeq 1.4556$ \cite{Salas}.

\section{Results for the 2A2B model}

\begin{figure}
\includegraphics[width=55mm,clip=]{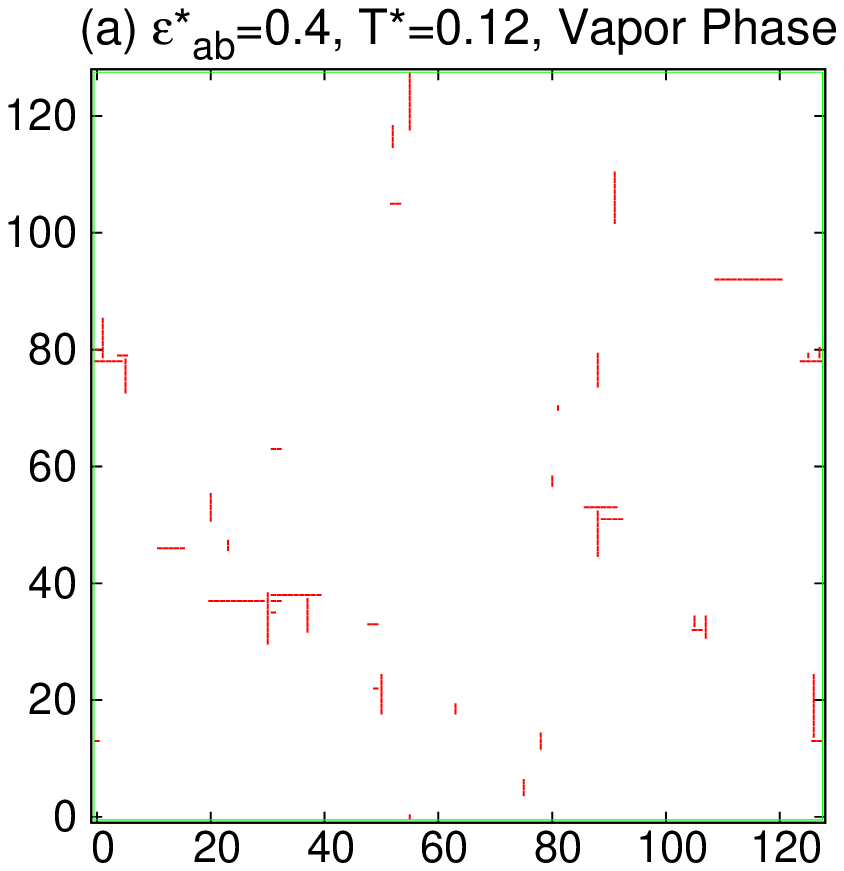}
\includegraphics[width=55mm,clip=]{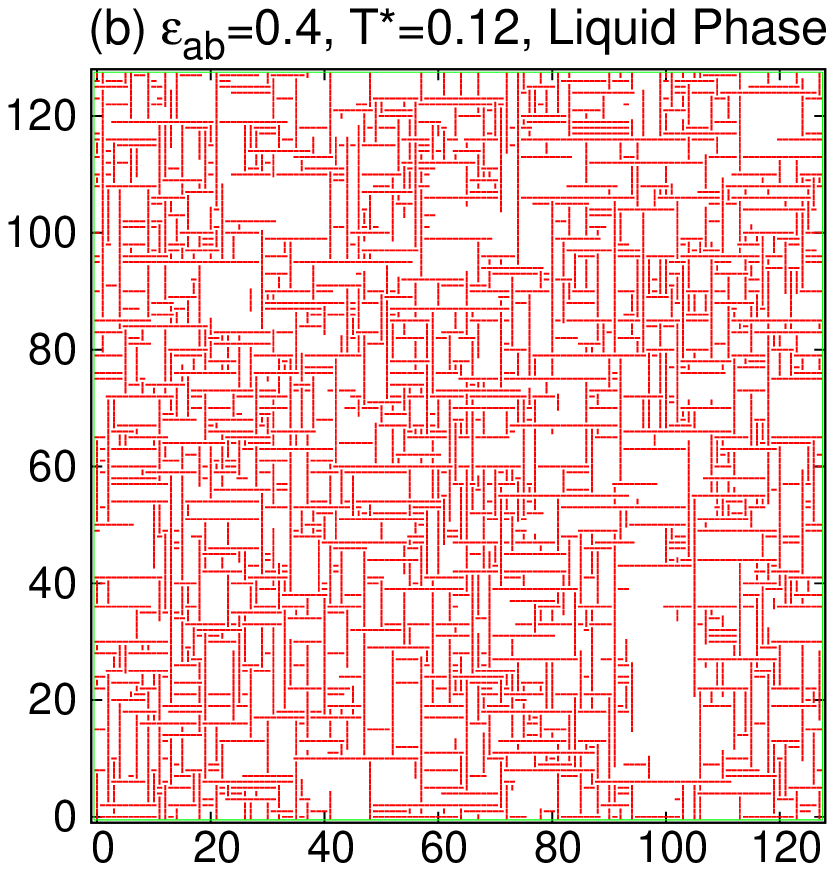}
\includegraphics[width=55mm,clip=]{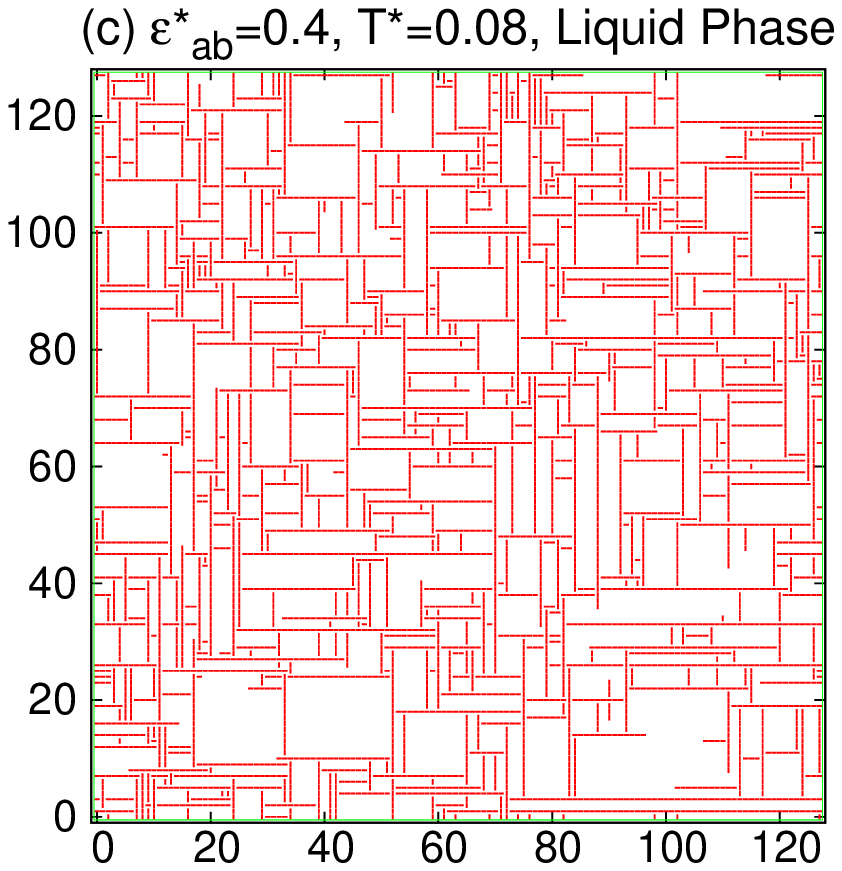}
\includegraphics[width=55mm,clip=]{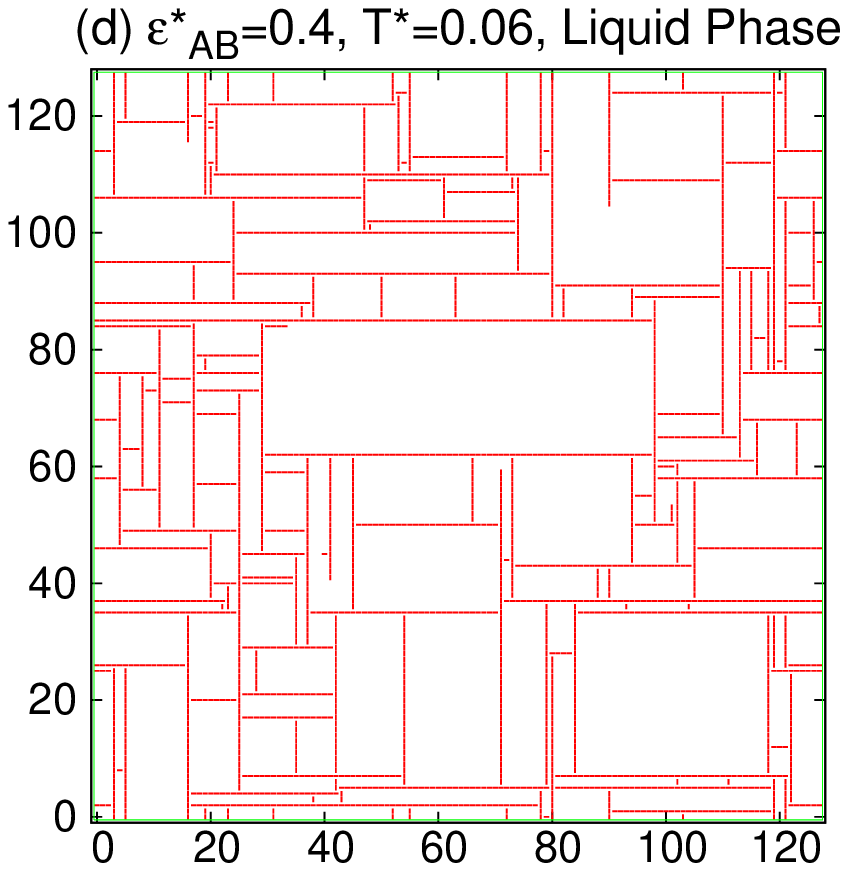}
\caption{Representative configurations of the $2A2B$ model with $\epsilon_{AB}^*=0.40$ and $L=128$ at liquid-vapor
coexistence, and several reduced temperatures $T^*=k_B T/\epsilon$. 
(a) Vapor phase at  $T^*=0.12$; (b) Liquid phase at $T^*=0.12$;
(c) Liquid phase at $T^*=0.08$; and (d) Liquid phase at $T^*=0.06$.
Particles are represented as segments of unit length oriented in the direction of the $A$ patches.
Note that the liquid becomes emptier as the temperature decreases. }
\label{Fig02}
\end{figure}

We start by illustrating, in Figure \ref{Fig02}, typical configurations of the coexisting phases for a system in the re-entrant regime, with $\epsilon_{AB}^*= \epsilon_{AB}/\epsilon_{AA} =0.40$, at three temperatures. We note that the density of the liquid decreases rapidly as the temperature decreases. The (network) liquid phase is characterized by voids (regions without particles) that increase as the temperature decreases. This observation 
implies that larger system sizes are required at lower temperatures, in order to sample adequately the increasing length scales that characterize the 
empty liquid phase.

\subsection{The phase diagram}

We consider $2A2B$ models characterized by different $\epsilon_{AB}^*$. After preliminary WLMC tests we choose appropriate subcritical temperatures
and compute the liquid-vapor equilibria by extrapolating to the thermodynamic limit the results of several system sizes. We then select a temperature 
(for each model) as the starting point of the GDI. These are collected in Table \ref{Table01}
\begin{table}
\begin{tabular}{|cc|lll|rr|}
\hline
$\epsilon_{AB}^*$ & $T_0^*$ & $\mu/\epsilon$ & $\rho_m$ & $\delta \rho$ & $L$(GDI-LT) & $L$(GDI-HT)	\\
\hline
0.375 & 0.10 &  -1.01582(2) & 0.155(1) & 0.146(3) & 512 & 512 \\
0.400 & 0.13 &  -1.04302(2) & 0.249(1) & 0.226(1) & 512 &  512 \\
0.450 & 0.15 &  -1.08487(2) & 0.3520(2)& 0.3376(2) & 512 & 256 \\
0.500 & 0.20 &  -1.16526(2) & 0.3891(4) & 0.3242(5) & 128 & 64 \\
0.550 & 0.22 &  -1.22628(2) & 0.4251(2) & 0.37544(3) & 128 & 128 \\
0.600 & 0.25 &  -1.30164(2) & 0.4426(3) & 0.3704(2) & 128 & 128 \\
\hline
\end{tabular}
\caption{Liquid vapor equilibria from WLMC simulations. 
The temperatures are those used as the starting points of subsequent GDIs. 
$L$(GDI-LT) and $L$(GDI-HT) correspond to the largest system sizes used in the GDI for temperatures 
below and above (respectively) the starting temperature $T_0$. Error bars, between parentheses, are given in
units of the last digit and correspond to a confidence level of about 95 \%.}
\label{Table01}
\end{table}

\begin{figure}
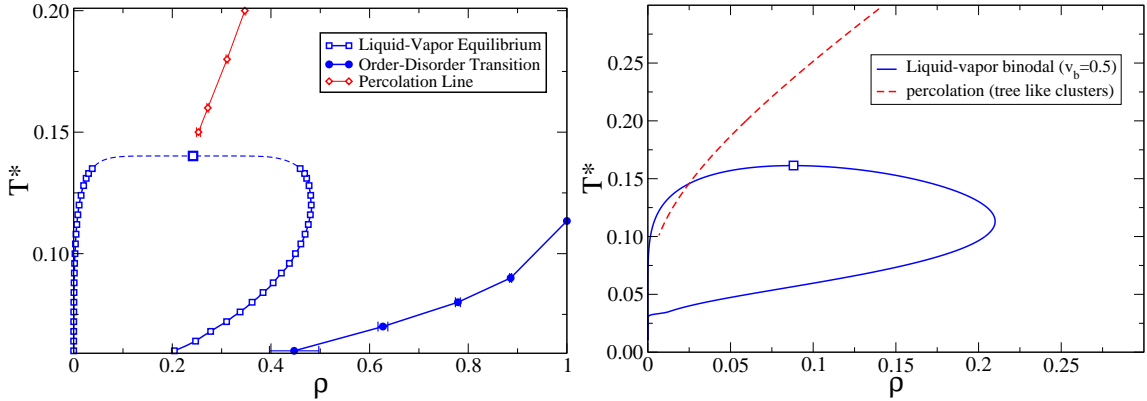

\includegraphics[width=75mm,clip=]{Fig03a.eps}
\includegraphics[width=75mm,clip=]{Fig03b.eps}
\caption{Phase diagram of the $2A2B$ model with $\epsilon_{AB}^*=0.40$. Left panel: Simulation results. Right panel: Results of  Wertheim's theory. See the legends for details.}
\label{Fig03}
\end{figure}

In Figure \ref{Fig03} we plot simulation and theoretical results for the liquid-vapor binodal of the $2A2B$ model with $\epsilon_{AB}^*=0.40$. The binodal has 
the ``pinched'' or re-entrant shape, characteristic of 3D off-lattice patchy particle models, 
with two $A$ patches and $\frac{1}{3}< \epsilon_{AB}^* < \frac{1}{2}$  \cite{[21],[22]}. The coexisting liquid density vanishes rapidly as the temperature decreases and the model exhibits an empty fluid regime. The theory (based 
on Wertheim's theory for associating fluids discussed in section V) describes the re-entrant behavior of the binodal and gives a good estimate of the critical temperature but underestimates the coexisting liquid density, as in related 3D off-lattice models \cite{[21],[22],[23],[24],[25]}. The computed percolation threshold, for clusters of particles connected by bonds between patches \cite{[22]}, is also shown in Figure \ref{Fig03}. 
The simulation results suggest that the percolation line intersects the LVE binodal at the critical point, in line with results for the 2D Ising model \cite{Coniglio_1977}. This contrasts 
with the results of Wertheim's theory (details of the theoretical methodology can be found in Ref. \onlinecite{[23]}) and the simulation results of 3D off-lattice models, where the percolation line intersects the LVE binodal on the vapor side \cite{[23]}.   

\subsection{The nature of the liquid-vapor critical points}

\begin{figure}
\includegraphics[width=120mm,clip=]{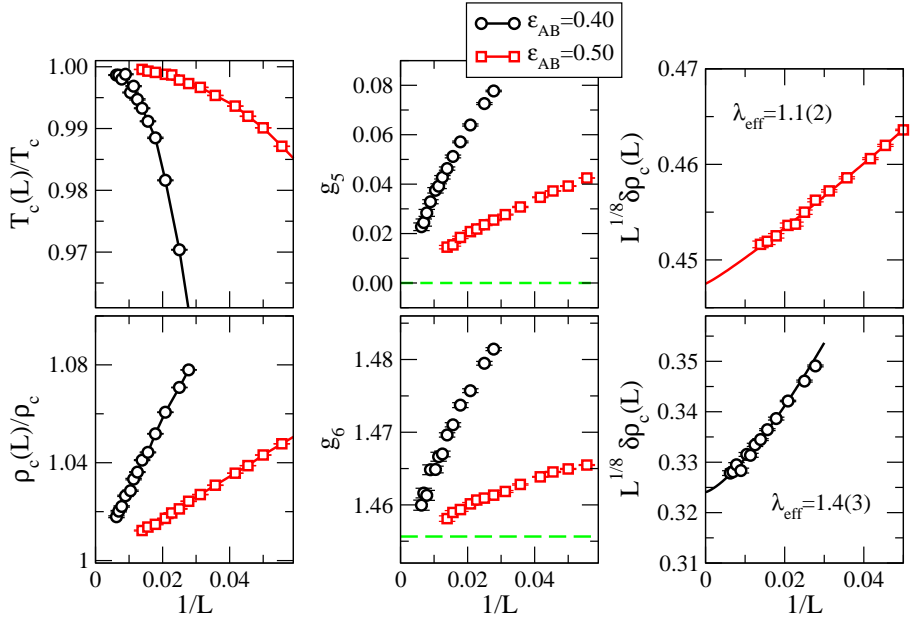}
\caption{Scaling of the critical parameters and of the moments of the density distribution function, at the pseudo-critical point, with the system size, for 
two $2A2B$ models; one model is in the re-entrant regime and the other is at the boundary to normal liquid behavior. The scaling results are fully consistent with 2D Ising criticality. 
The pseudocritical parameters, $\rho_c(L)$, $T_c(L)$ and $\delta \rho_c(L)$, follow the 2D Ising scaling laws, Eqs. (\ref{eq.rhoc}), (\ref{eq.tc}) and (\ref{eq.ddc});  while the moment ratios, $g_5$, and $g_6$, approach the 2D Ising values as $L$ increases. It is also clear that the scaling region decreases as $\epsilon_{AB}^*$ decreases.}
\label{Fig04}
\end{figure}

In Figure \ref{Fig04} we illustrate the scaling behavior of the critical parameters and the moments of the density distribution function, at the LVE pseudo-critical points, with the system size, for two $2A2B$ models with: $\epsilon_{AB}^*=0.40$ and $\epsilon_{AB}^*=0.50$. The observed behavior is consistent with criticality in the 2D Ising universality class. Note, however, that the system size dependence of the critical properties is stronger in the system with $\epsilon_{AB}^*=0.40$. 

The results for the critical temperature and the critical chemical potential hardly depend on whether we use $\lambda=4/3$;   $\lambda=7/4$ or $\lambda$ being considered as a fitting parameter (See Eq.\ref{eq.tc}).
In the latest case the effective values of $\lambda$ are always larger than $\lambda=1$ (See the effective values $\lambda_{eff}$ in Table \ref{Table02}). In particular, for the largest values of $\epsilon_{AB}^*$ the effective values of $\lambda$ are consistent with $\lambda=4/3$. For $\epsilon_{AB}^* = 0.35, 0.40$ the uncertainty in the effective value of $\lambda$ is too large to discriminate between the two scaling scenarios. Nevertheless, the fact that the effective values of $\lambda$ satisfy $\lambda > 1$, supports the hypothesis that the LVE critical point of the $2A2B$ model is in the 2D universality class. 

\begin{figure}
\includegraphics[width=90mm,clip=]{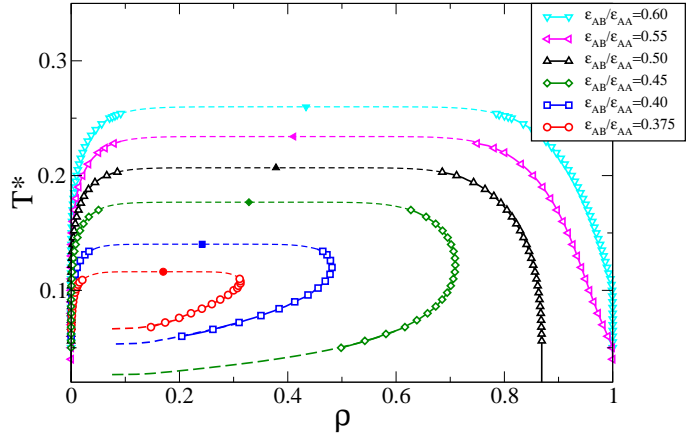}
\caption{Simulation results for the liquid-vapor binodals of different $2A2B$ models 
in the re-entrant and normal regimes. From top to bottom:
$\epsilon_{AB}^* = 0.60, 0.55, 0.50, 0.45, 0.40, 0.375$.  }
\label{Fig05}
\end{figure}

In Figure \ref{Fig05} we plot the liquid-vapor phase diagram for various $2A2B$ models in the re-entrant and normal regimes. Numerical results for the 
critical points are collected in Table \ref{Table02}. The results shown for $T_c$ and $\mu_c$ are
those extracted from the fitting scheme with fixed $\lambda$ that provides the best agreement
with simulation data ($\lambda=7/4$ for $\epsilon_{AB}^* \le 0.40$, and $\lambda=4/3$ for 
$\epsilon_{AB}^* > 0.40$).
As expected both the critical temperature and the critical density increase with $\epsilon_{AB}^*$. A significant change in the binodal, however, occurs 
at $\epsilon_{AB}^*=0.5$. For models with $\epsilon_{AB}^* > 0.5$ the liquid density approaches $\rho=1$ as the temperature vanishes; while for models with $\epsilon_{AB}^* < 0.5$ the liquid density decreases with temperature, at low temperatures and seems to approach $\rho=0$ as the temperature vanishes.
This conclusion is based on theoretical results (see section V) and confirmed by computer simulations of 3D off-lattice models \cite{[14],[15],[21],[22]}. 
The simulation of systems at vanishingly low temperatures is hindered by two factors: the usual problems of sampling at low temperatures, and the emergence 
of diverging length scales, namely the size of the voids in the empty liquid phase.

The model with $\epsilon_{AB}^*=0.50$ exhibits an intermediate behavior. Simulation results suggest that at $T=0$ the density of the liquid phase at 
equilibrium with the vapor approaches a finite density, $\rho \approx 0.87$. Note also the dashed lines that continue the liquid branches of the models 
with $\epsilon_{AB}^*<0.50$. These lines were computed using a related model that captures the phase behavior of the $2A2B$ patchy particle models at low 
temperatures, to be described in section IV.
\begin{table}
\begin{tabular}{|c|llll|rr|}
\hline
$\epsilon_{AB}^*$ & $T_c^*$ & $\mu^*_c$ & $\rho_c$ & $\lambda_{\textrm eff}$  & $L_{\min}$ & $L_{\max}$ \\
\hline
0.375 &  0.1160(10) & -1.0251(6) & 0.170(5) & 2.3$\pm$ 1.6 & 64 &128 \\
0.400 & 0.1402(2) & -1.0509(2)  & 0.242(1) & 1.7$\pm$ 0.3 & 36 & 160 \\
0.450 & 0.1768(2) & -1.1087(2)  & 0.3279(5) & 1.3 $\pm$ 0.3 & 20 &64 \\
0.500 & 0.2068(1) & -1.1715(1)  & 0.3782(3) & 1.3 $\pm$ 0.2 & 12 &72 \\
0.550 & 0.2340(1) & -1.2387(1)  & 0.4109(4) & 1.2 $\pm$ 0.3 &10 &56 \\
0.600 & 0.2599(2) & -1.3099(2)  & 0.4339(3) & 1.3 $\pm$ 0.2 & 8  & 32 \\
\hline
\end{tabular}
\caption{Numerical results for the liquid-vapor critical points of different $2A2B$ models in the re-entrant and normal regimes.
The effective exponents $\lambda_{\textrm eff}$ are obtained from the fits of the pseudocritical temperatures using
Eq. (\ref{eq.tc}) with $\nu=1$. Error bars correspond to a confidence level of about 95\%. In
the cases of $T_c$ and $\mu_c$ the error bars extend over the results of the three fitting
schemes ($\lambda=4/3$, $\lambda=7/4$; and $\lambda$ as adjustable parameter).  }
\label{Table02}
\end{table}

A final important question remains. The $2A2B$ patchy particle models undergo, in general, two thermodynamic transitions, a first-order liquid-vapor transition at low densities; and a continuous O-D transition at high densities (See the phase diagram for $\epsilon_{AB}^*=0.40$ in Figure \ref{Fig03}). Previous results \cite{[21],[22]} and those discussed here suggest that for appropriate values of $\epsilon_{AB}^*$ both the vapor and the liquid branches of 
the LVE approach zero density at zero temperature; on the other hand the results for the SARR model \cite{[31],[32],[33]} also indicate that the O-D transition approaches zero density at zero temperature.

Several scenarios are then plausible for the global phase diagram of the $2A2B$ patchy particle model depending on where the O-D
critical line intersects the LVE line; 
this can happen at a finite temperature (either at a critical end point or at a tricritical point) or at $T=0$. The simulation 
results for $\epsilon_{AB}^*=0.45$, and $\epsilon_{AB}^*=0.40$, discard the possibility of an upper tricritical point, as the critical temperature of the LVE is 
higher than the exact result for the O-D transition at $\rho=1$ and previous \cite{[32]} and present results (See Figure \ref{Fig03}) suggest that the critical temperature of the O-D transition increases with the density. 
 
The question remains whether the transitions meet at $T=0$ or at a finite temperature critical end point. For the models with $\epsilon_{AB}^*<1/2$ considered here we computed the order parameter on the liquid branch as obtained from the GDI, for several system sizes, and found that the network liquid phase at LVE is orientationally disordered. If a critical end point exists then it has to occur at lower temperatures than those accessible by simulations of the $2A2B$ patchy particle model. To proceed we consider a related model in the next section.

\section{Phase behavior at low temperatures}

Let us consider $2A2B$ models with $\epsilon_{BB}=0$. Defining $\lambda = \frac{1}{2} -\epsilon_{AB}^*$ we re-write Eq. (\ref{eq3}) as:
\begin{equation}
{\cal U}/\epsilon = - N + \lambda {\cal N}_{AB}  
 + \frac{1}{2} {\cal N}_{A0}, \
\end{equation}
with $\lambda=0$ for $\epsilon_{AB}^* = \epsilon/2$, and $\lambda=1/2$ for $\epsilon_{AB}^*=0$. Now consider $0 \le \lambda << 1/2$. At sufficiently 
low temperatures ${\cal N}_{A0}$ (the number of non-bonded $A$ patches)  is negligible with respect to both ${\cal N}_{AB}$ and $\lambda {\cal N}_{AB}$, and the thermodynamics of the model 
is determined by $\lambda^* \equiv \lambda \epsilon/k_B T$. This suggests that the LVE of $2A2B$ patchy particle models at low temperatures, and small 
$\lambda$, may be collapsed (approximately) onto a single curve.

\subsection{The Low Temperature Model}

The previous discussion suggests the consideration of a related low temperature model (LTM) with interaction energies: $u_{A0}= \infty$ (i.e. non-bonded 
$A$ patches are disallowed), $u_{AA}=u_{BB}=u_{B0}=0$, and $u_{AB}=\lambda \epsilon$. 

In order to compute the LVE we assume that the vapor phase at low temperatures has zero density. This results from the fact that, as all $A$ patches 
are bonded, the particles must belong to a network that percolates in, at least, one direction; since the vapor does not percolate its density must 
vanish in the thermodynamic limit.
We compute the liquid branch of the LTM using Gibbs-Duhem integration, with thermodynamic variables $\lambda^*$ and $(\beta \mu) \equiv \mu/k_BT$. The differential equation to be solved is:\cite{Almarza-Noya}
\begin{equation}
N d (\beta \mu) - {\cal N}_{AB} d \lambda^* = 0
\end{equation}
\begin{equation}
\left( \frac{  \partial (\beta \mu) }{\partial \lambda^* } \right)_{\textrm coex} =  \frac{< {\cal N}_{AB} > }{< N >  }
\end{equation}
where $N$ and ${\cal N}_{AB}$ correspond to the liquid branch (the vapor has zero density). The starting point for the integration is $(\beta \mu)=(\beta\mu)_0$, $\lambda^*=0$. The calculation of $(\beta \mu)_0$ is discussed in the Appendix.
At full lattice occupancy the LTM is equivalent to the original patchy particle model, since all $A$ patches are bonded, and thus the LTM may also be used to compute the O-D transition. 

In the LTM every $A$ patch is bonded (either to another $A$ or to a $B$ patch). In addition, the density of the liquid phase decreases rapidly as $\lambda^*$ increases, in the empty fluid regime. A standard algorithm involving single particle moves is useless under these conditions. In order to sample the LTM we develop an efficient cluster algorithm that is described below.

\subsection{The cluster algorithm}
The LTM algorithm is based on three types of moves:
(a) Rotation of particles (only if the four NNs of the particle are occupied);
(b) Insertion of a sequence of aligned particles;
(c) Deletion of a sequence of aligned particles.

The rotation move is straightforward. One particle with four NNs occupied is selected at random (if there is any), and then one of its two orientations is 
chosen with a probability proportional to its Boltzmann factor.

The insertion / deletion of a sequence of aligned particles is carried out as follows:
A lattice site is chosen at random: If the site is occupied then a deletion attempt is performed. It starts by identifying the linear cluster of particles 
linked to the selected one by an unbroken sequence of $AA$ bonds. Such cluster either percolates through the periodic boundary conditions (PBC) or ends at 
two $AB$ bonds. If the removal of the cluster leads to an unbonded $A$ patch the deletion attempt is rejected, otherwise the acceptance criterion (defined 
below) is applied.

If the chosen site is empty then one direction, $s=1,2$, is chosen at random. A linear $AA$ cluster of occupied sites is built along the chosen direction 
(on both sides), the bonding criterion being that the NN position is empty. The process stops when the cluster percolates through the PBC or when occupied 
sites are found at both ends. The acceptance criterion (defined below) is then applied.

It is straightforward to compute the change in energy when inserting or deleting a LTM cluster. The cluster either percolates through a sequence of 
$AA$ bonds ($\Delta U^* = 0$) or terminates at both ends with $AB$ bonds: $\Delta U= \pm 2\lambda$. Considering that positions (not insertions/deletions) 
are selected at random, the acceptance probabilities are:
\begin{equation}
\frac{ A(N+\Delta N|N) }{ A(N|N+\Delta N)} = 2 \: \exp \left[ - \beta \Delta U + \beta \mu \Delta N \right]
\end{equation}
where $\Delta U = U_{N+\Delta N} - U_N$, and the factor $2$ arises from the two orientations of the inserted cluster, of length $\Delta N$ lattice sites.

\subsection{Simulation Results}

The GDI requires as input a point on the LVE binodal, which was taken to be $(\beta \mu)_0$, the reduced chemical potential at $\lambda^*=0$. 
The chemical potential at zero pressure (the vapor phase has zero density) is obtained via thermodynamic integration \cite{frenkel-smit} from ($\beta \mu \rightarrow \infty$), as the partition function in the full lattice limit and $\lambda=0$ is known exactly: $Q = 2^N$. We carried out the calculation for different system sizes $L=16, 32, 64, \cdots, 256$ and found that the size dependence of $\mu$ is negligible, obtaining $(\beta \mu)_0 = -0.78940(2)$. This 
is consistent with the estimate from the GDI of the patchy particle model with $\epsilon_{AB}^*=0.50$, which gives $\mu/\epsilon -1 \simeq  - 0.789 T^*$, 
at low temperatures.
\begin{figure}
\includegraphics[width=120mm,clip=]{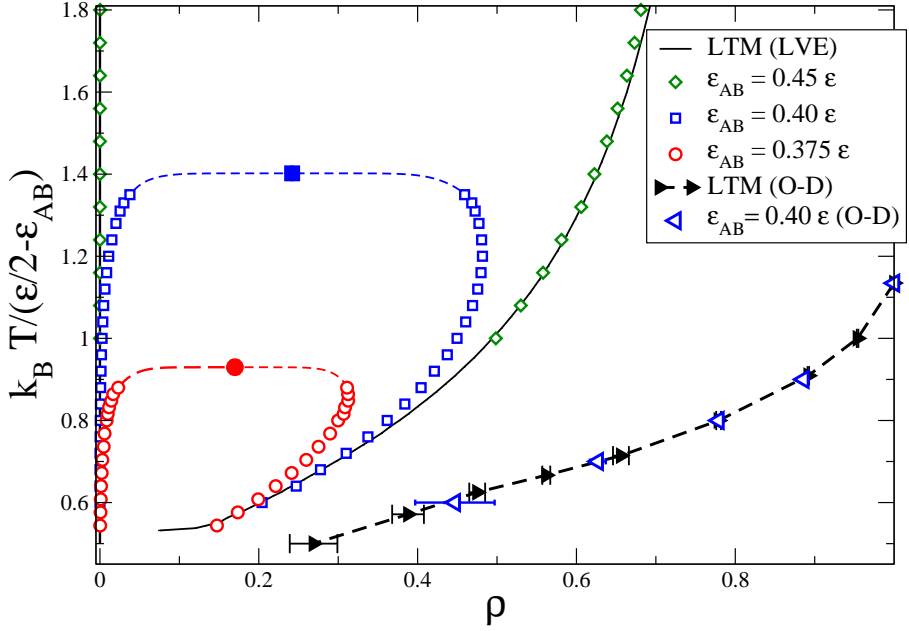}
\caption{Phase diagram of the LTM (LVE and order-disorder transition);
re-scaled liquid-vapor binodals of the $2A2B$ patchy particle models with
$\epsilon_{AB}^*<0.50$, and order-disorder transition for $\epsilon_{AB}^*=0.40$ (Symbols
are explained in the legends).
The liquid branch of the LTM model is computed for a system with $L=2048$.}
\label{Fig06}
\end{figure}
The LVE is obtained using Gibbs-Duhem integration. Several system sizes are considered to check the system size dependence of the results. 
In Figure \ref{Fig06} we test the accuracy of the LTM to describe the coexisting liquid densities of the $2A2B$ patchy particle models, at low temperatures. 
Clearly, the LTM results converge to those of the patchy particle models as the temperature is lowered. 
We found that as $\lambda^*$ increases (scaled 
temperature decreases) larger systems are required to obtain consistent results for different system sizes (as was observed in the simulation of the 
$2A2B$ patchy particle models). Indeed, the line corresponding to the LTM liquid branch in Figure \ref{Fig06} is plotted for scaled temperatures higher 
than those where the results for systems with $L=1024$ and $L=2048$ start to show significant differences: 
i.e. $\lambda^* \simeq 1.88$. This is due 
to the rapid increase of the size of the voids in the empty liquid at low temperatures, which hinders the simulations of the LTM at larger $\lambda^*$ 
due to the system size requirements and the loss of efficiency of the simulation algorithm.

Finally, we have computed the O-D transition of the LTM, which is almost indistinguishable from that of the $2A2B$ model with $\epsilon_{AB}^*=0.40$, after proper re-scaling (both are plotted in Fig. \ref{Fig06}).

Despite the difficulties in simulating the LVE of the LTM when $\lambda^* \ge 1.9 $, 
the numerical results suggest that the ratio $\rho_{L}(\lambda^*)/\rho_{OD}(\lambda^*)$ decreases with $\lambda^*$, 
for $\lambda^* \gtrsim  1.60 $. Considering that the LTM describes accurately the low temperature phase diagram of the $2A2B$ patchy particle models, we 
conclude that the most likely topology of the phase diagram of 
this class of models is characterized by a multicritical point at $T=0$ and $\rho=0$, where the liquid-vapor and the O-D transitions merge.

\section{Topology of the phase diagram: Theory}

\subsection{Wertheim's Theory}

In this section we address the topology of the phase diagram by resorting to theoretical/analytical techniques. The thermodynamics of the $2A2B$ patchy particle model can be described using Wertheim's first order perturbation theory (WPT), which accounts accurately, in the low density limit, for the effect of association 
\cite{[21],[22]}. The reference free energy $F_{ref}$ is that of an ideal lattice gas, 
\begin{equation}
\label{freenid}
\frac{\beta F_{ref}}{N}=
\ln \rho +\frac{1-\rho}{\rho}\ln(1-\rho),
\end{equation}
where $\rho$ is the density. The perturbation term $F_b$ includes the bonding contribution and is given, within WPT, by \cite{[21]}
\begin{equation}
\label{werthiso}
\frac{\beta F_b}{N}= 2\ln X_A-X_A +2\ln X_B -X_B,
\end{equation}
where $X_\alpha$ is the probability that a bonding site of type $\alpha$ is unbonded. These probabilities are related to the thermodynamic quantities 
through the laws of mass action (i.e. by considering bond formation as an equilibrium chemical reaction), which are, for particles with $2A$ 
and $2B$ bonding sites,
\begin{equation}
\label{lmaXalph}
X_\alpha+2\rho\Delta_{\alpha\alpha}X_\alpha^2+
2\rho\Delta_{\alpha\beta}X_\alpha X_\beta=1,
\end{equation}
with $\alpha=A,B $ and $\beta \ne \alpha$. The quantities $\Delta_{\alpha\beta}$ are given by, 
\begin{equation}
\label{Deltaij}
\Delta_{\alpha\beta}=v_{\alpha\beta}
\left[\exp(\beta \epsilon_{\alpha\beta})-1\right],
\end{equation}
with $v_{\alpha\beta}$ the volume of the $\alpha\beta$ bond (in units of the volume of a lattice site). Note that, since $\epsilon_{BB}=0$, $\Delta_{BB}=0$. We take $v_b=1/2$, in order to maximize $v_b$ while disallowing more than one bond between 
two patches and more than two patches per bond (see Figure \ref{Fig01}).

\begin{figure} 
\includegraphics[width=80mm,clip=]{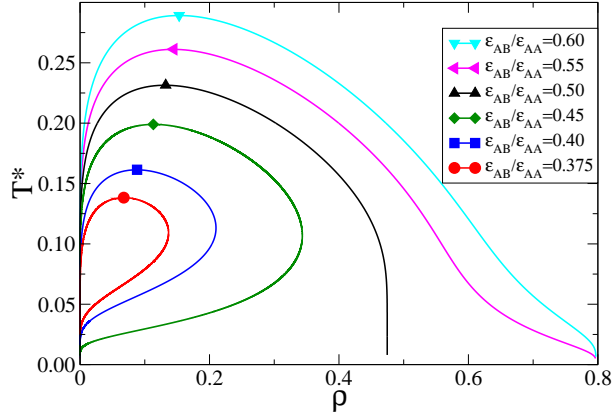}
\caption{Phase diagram of the $2A2B$ model, based on Wertheim's theory, for several values of $\epsilon_{AB}^*$.}
\label{Fig07}
\end{figure}

Using equations (\ref{lmaXalph}) in (\ref{werthiso}), $F_b$ is obtained as a function of $\rho$ and $T$. From the Helmholtz free energy $F=F_{ref}+F_b$, 
one can obtain the pressure and the chemical potential and calculate the phase diagram. 
Figure \ref{Fig07} shows the results of this calculation for several values of $\epsilon_{AB}^*$. Comparison with the results of simulations (section III) 
reveals that the theory describes correctly the re-entrance of the liquid branch for $\epsilon_{AB}<0.5 \epsilon$ and the constant density of the liquid 
branch at low temperatures when $\epsilon_{AB}=0.5 \epsilon$.
The theory also predicts \cite{[21],[14]} that no liquid-vapor coexistence occurs when $\epsilon_{AB}<\epsilon/3$, in line with the results of the 
simulations for these values of the parameters.  
Note that, as in 3D off-lattice models \cite{[21],[23]}, there is almost quantitative agreement between the critical temperatures obtained by theory and simulations, while the theory underestimates systematically the density of the coexisting liquid branch.

\subsection{The liquid branch of the binodal and the order-disorder transition}

Wertheim's theory, as described in the previous section, is not capable of describing the O-D transition. The ordering is driven  by the 
excluded volume of the chains formed at low temperatures \cite{[31]}, and this effect is not included in (\ref{freenid}) nor in (\ref{werthiso}). Based on previous works \cite{[31],[21]}, we proceed to derive asymptotic expressions for the O-D transition and the liquid branch of the binodal. This analysis 
shows that the line of LVE is not intersected by the O-D line at any finite temperature, and thus a critical end point does not occur in this model.

\begin{figure} 
\includegraphics[width=80mm,clip=]{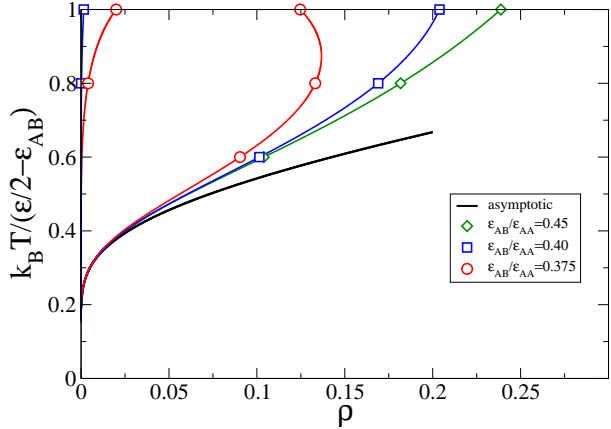}
\caption{Binodals for $\epsilon_{AB}^*<0.5$ at low densities and temperatures calculated using Wertheim's theory. The black line is the asymptotic result 
(\ref{binliquidass}) for the liquid branch of the binodal. Note that the temperature of each binodal is rescaled by $\lambda=1/2-\epsilon_{AB}^*$.}
\label{Fig08}
\end{figure}

The asymptotic limit for the liquid branch of the binodal of $2A2B$ models in the re-entrant regime, $\epsilon_{AB}<0.5\epsilon$, is obtained using the 
results of \cite{[21]}. Taking into account that the reference free energy is given by (\ref{freenid}), the asymptotic pressure, i.e. the pressure in 
the limit of strong $AA$ association within WPT, is given by,
\begin{equation}
\label{pressass}
\beta p =a_0\rho^{\frac{1}{2}}-a_1\rho^{\frac{3}{2}}+\frac{\rho^2}{2},
\end{equation}
with $a_0=(2\Delta_{AA})^{-\frac{1}{2}}$ and $a_1=2\Delta_{AB}a_0$. The first term vanishes when all the $A$ patches are bonded (see section IV) and under 
these conditions $p\approx 0$ at coexistence. The coexisting liquid density, $\rho_\ell$, is then approximated, at low densities and temperatures, by, 
\begin{equation}
\label{binliquidass}
\frac{k_B T}{\epsilon_{AA}\lambda}=-\frac{2}{\ln\rho_\ell-2\ln2}.
\end{equation}
The asymptotic liquid density is plotted in Figure \ref{Fig08} together with the binodals of various $2A2B$ models, obtained using Wertheim's theory. 
It is clear that Eq.(\ref{binliquidass}) is the asymptotic limit of the liquid branch of $2A2B$ patchy particle binodals, at low densities and temperatures.

\begin{figure} 
\includegraphics[width=80mm,clip=]{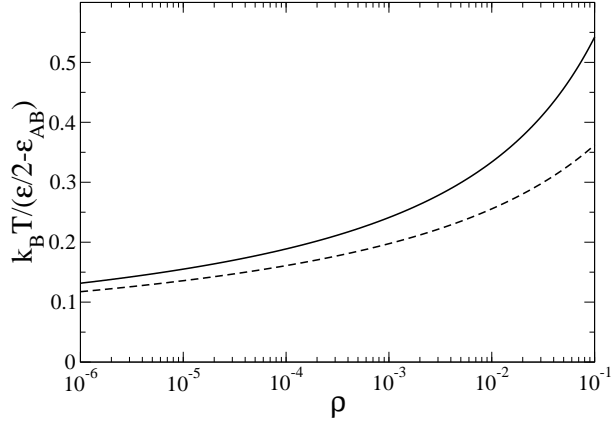}
\caption{Asymptotic results for the liquid branch of the binodal 
(full line) given by (\ref{binliquidass}) and 
for the order disorder transition (dashed line) given by (\ref{orddisass}).}
\label{Fig09}
\end{figure}

Finally, we turn our attention to the O-D transition of the $2A2B$ patchy particle models. In \cite{[31]} the SARR model, which is 
the limit of the $2A2B$ model when $\epsilon_{AB}=0$ (i.e. when only $AA$ chains are formed), was investigated and the contribution of the excluded volume 
of two chains was included in the free energy via an Onsager like approximation. 
An order parameter $\Delta=\rho_x-\rho_y$ (with $\rho_{\alpha}$ being the number density of particles with $A$
patches aligned along $\hat \alpha$) was defined, and it was shown that the field $h_0$ associated to $\Delta$, is,
\begin{equation}
\label{hfield0}
\beta h_0= \frac{1}{2}\left[\ln\left(\frac{X}{Y}\right)-\Delta\right],
\end{equation}
with $X$ and $Y$ given by,
\begin{equation}
\rho_x=\exp(-\beta\epsilon)\frac{X}{(1-X)^2},
\end{equation}
\begin{equation}
\rho_y=\exp(-\beta\epsilon)\frac{Y}{(1-Y)^2}.
\end{equation}

The O-D line is found by solving 
$\left(\frac{\partial h_0}{\partial \Delta}\right)_{\Delta=0}=0$.

The $2A2B$ model differs from the SARR model by allowing the formation of $AB$ bonds. In the limit of strong $AA$ bonds the field $h$, conjugated to the order 
parameter $\Delta$, may be taken to be,
\begin{equation}
\label{hfield}
\beta h= \beta h_0 +\left(\frac{\partial \beta f_{AB}}{\partial \Delta}\right),
\end{equation}
where $f_{AB}$ is the contribution of the $AB$ bonds to the free energy density. If this free energy is calculated within WPT and taken in the limit 
of strong $AA$ bonds \cite{[21]},
\begin{equation}
\beta f_{AB}=-4\sqrt{2}a_1\left(\rho_x^{\frac{1}{2}}\rho_y+
\rho_y^{\frac{1}{2}}\rho_x\right),
\end{equation}
one obtains a simple expression for $h$,
\begin{equation}
\beta h= \beta h_0-2
\sqrt{2} a_1
\left(\frac{1}{2}(\rho_x^{-\frac{1}{2}}\rho_y +
\rho_y^{-\frac{1}{2}}\rho_x)+ 
\rho_x^{\frac{1}{2}}+\rho_y^{\frac{1}{2}}\right). 
\end{equation}
The O-D line is now calculated by solving $\left(\frac{\partial h}{\partial \Delta}\right)_{\Delta=0}=0$, which, in the limit of low densities and temperatures, yields,
\begin{equation}
\label{odasymp}
a_0\rho^{\frac{1}{2}}-\rho^2+5a_1\rho^{\frac{3}{2}}=0.
\end{equation} 
In line with the derivation of (\ref{binliquidass}), we neglect the first term of Eq. (\ref{odasymp}), and obtain for the O-D transition line,
\begin{equation}
\label{orddisass}
\frac{k_B T}{\epsilon_{AA}\lambda}=-\frac{2}{\ln\rho-5\ln2}.
\end{equation}
The asymptotic liquid binodal (\ref{binliquidass}) and the asymptotic O-D line (\ref{orddisass}) are plotted in Figure \ref{Fig09}:
they do not intersect at finite temperature and thus the global phase diagram of the $2A2B$ model does not have a critical end point. In other words, the empty liquid regime is not pre-empted by the ordered (liquid) phase. 

\section{Conclusions}

We investigated a simple patchy particle lattice model consisting of particles with four bonding sites, two of type $A$ and two of type $B$, on the square lattice, and computed its global phase behavior by simulations and theory. We have set the interaction between $B$ patches to zero and calculated the phase diagram as the ratio between the $AB$ and the $AA$ interactions, $\epsilon_{AB}^*$, varies. In line with previous work on three-dimensional off-lattice models, we have shown that the liquid-vapor phase diagram exhibits a re-entrant or ``pinched'' shape, for an identical range of $\epsilon_{AB}^*$, suggesting somewhat surprisingly that this range - and the corresponding empty fluid regime - is independent of the dimensionality of the system and of the lattice structure.

In addition, the use of low-dimensional lattice systems allowed the simulations of much larger systems enabling us to establish the nature of the liquid-vapor critical points, which were found to be in the Ising 2D class, both in the re-entrant and the normal liquid models. While in the normal liquid regime the 
scaling regions are typical of models in the 2D Ising universality class, in the re-entrant liquid regime the scaling region decreases rapidly as the
critical temperature (or $\epsilon_{AB}^*$) decreases. Our theoretical and simulation results also suggest that the Ising scaling region vanishes as the critical temperature vanishes, in line with the presence of a multicritical point at zero density and temperature. 

The patchy particle models on the square lattice exhibit an O-D transition at fixed density, as the temperature is lowered below the bonding temperature. This transition is anti-ferromagnetic for normal liquid models and ferromagnetic for models with a re-entrant liquid regime. In the latter models, the results of simulations of an appropriate low-temperature-model that describes the asymptotics of the particle patchy systems at low temperatures, together with an efficient sampling cluster algorithm, suggest that the line of O-D transitions intersects the condensation line at zero temperature and zero density.
This topology of the phase diagram is corroborated by an asymptotic theoretical analysis of the liquid branch of the binodal and of the O-D transition, based on Wertheim's theory for associating fluids. The theory is exact at zero density, lending support to the results of the asymptotic analysis in the low temperature, low density region. 

In summary, the results of simulations and of theoretical analysis strongly suggest that the line of O-D transitions intersects the condensation line at a multicritical point at zero temperature and density, for patchy particle lattice models in the re-entrant liquid regime. The global phase diagram of off-lattice patchy particle models, in 2D and 3D, is further complicated by the presence of stable solid phases. These phases may pre-empt the empty fluid regime rendering the zero temperature zero density multicritical point metastable.

\acknowledgments
NGA gratefully acknowledges the support from the Direcci\'on General de Investigaci\'on 
Cient\'{\i}fica  y T\'ecnica under Grants No. FIS2010-15502, and from the
Direcci\'on General de Universidades e Investigaci\'on de la Comunidad
de Madrid under Grant No. S2009/ESP-1691 and Program MODELICO-CM. MMTG, JMT and MS acknowledge 
financial support from the Portuguese Foundation for Science and Technology (FCT) under 
Contracts nos. PEst-OE/FIS/UI0618/2011 and PTDC/FIS/098254/2008.

\section{Appendix: Computing the initial point for the GDI of the LTM}

The limit $\lambda^*=0$ of the LTM is an athermal model, where all allowed configurations have zero potential energy. A configuration is allowed 
if (and only if) every patch of type $A$ is bonded. In this model a first-order transition, corresponding to the transition of the $2A2B$ 
model with $\epsilon_{AB}=\epsilon/2$ at $T=0$, is expected to occur. The results of a series of simulations of the LTM with increasing / decreasing 
values of $\beta \mu$ indicate that the transition at $\lambda =0$ is indeed first order. The coexisting vapor phase is found to have vanishingly 
small density, in line with the results for the $2A2B$ model.

The value of $\beta \mu$ at the transition is computed using thermodynamic integration, as the partition function for $\lambda^*=0$ in the full 
lattice limit is known exactly:
\begin{equation}
Q(N=M,M) = q^M ,
\end{equation}
where $q$ is the number of particle orientations, $q=2$ for the square lattice. If the number of vacancies is small: $M-N << M$, we can write an 
approximate expression for the partition function, by assuming that the number of NN pairs of vacancies is negligible:
\begin{equation}
Q(N,M) \simeq Q_0(N,M) = 
2^{M-5(M-N)} \left( \begin{array}{c} M \\ N \end{array} \right)
=   2^{5N-4M} \left( \begin{array}{c} M \\ N \end{array} \right).
\end{equation}
The factor $2^{-5(M-N)}$ arises as an isolated vacancy eliminates the possibility of having two different states (orientations) at the vacant site and 
at its four NNs (that are assumed to be occupied), since in the LTM orientations with $A$ patches pointing to the vacant site are not allowed.

We can assume the vapor phase density to vanish, and thus the pressure at coexistence also vanishes. As the partition function at full coverage is known, 
we can compute the equation of state of the high density phase and the value of the chemical potential at the transition $\mu_{lv}$, which satisfies: 
$\beta p (\beta \mu_{lv}) = 0 $.

In order to derive a procedure to compute $\beta \mu_{lv}$ we consider, first, the approximate partition function $Q_0(N,M)$. The corresponding grand 
canonical partition function is:
\begin{equation}
Q_0(\beta \mu,M) ) =
  2^{-4M} 
 \sum_{N=0}^M 
\left( \begin{array}{c} M \\ N \end{array} \right)
  \exp \left[ \left( \beta \mu + 5 \ln 2 \right) N \right];
\end{equation}
which can be summed to give:
\begin{equation}
Q_0(\beta \mu,M) = 2^{-4M} 
\left[ 1 + \exp \left( \beta \mu'  \right) \right]^M;
\end{equation}
where $\beta \mu' = \beta \mu + 5 \ln 2 $.
The pressure, at this level of approximation, is written: 
\begin{equation}
\beta p^{(0)}(\beta \mu',M)  = \frac{1}{M} \ln Q_0(\beta \mu, M ) 
= - 4 \ln 2 + \ln \left( 1 + e^{\beta \mu'} \right); 
\end{equation}
At the same level of approximation, the density is easily computed:
\begin{equation}
\rho^{(0)}(\beta \mu',M) =\frac{1}{M} 
\frac { \sum_{N=1}^M 
\left( \begin{array}{c} M \\ N \end{array} \right) e^{\beta \mu' N} N }
{ \sum_{N=1}^M 
\left( \begin{array}{c} M \\ N \end{array} \right) e^{\beta \mu' N}  }
= 
\frac{ e^{\beta \mu' }} { 1 + e^{\beta \mu' } }
\end{equation}
We now define the fugacity fraction $\phi$ as:
\begin{equation}
\phi = \frac{e^{\beta \mu' } }{1+e^{\beta \mu'} },
\end{equation}
we obtain,
\begin{equation}
\rho^{(0)} (\phi) = \phi,
\end{equation}
\begin{equation}
\beta p^{(0)} (\phi) = - 4 \ln 2 - \ln \left( 1 - \phi \right) .
\end{equation}
In the Grand Canonical Ensemble for processes at constant temperature and constant volume, we have:
\begin{equation}
d (\beta p ) =   \rho d (\beta \mu),
\end{equation}
which can be integrated to give the pressure as a function of the fugacity fraction:
\begin{equation}
\beta p( \phi) =  \beta p^{(0)} (\phi) + \int_{1}^{\phi}
{\textrm d} \phi_1 \left[ \rho(\phi_1) - \phi_1 \right] 
\frac{d (\beta \mu) }{d\phi_1}, 
\end{equation}
\begin{equation}
\beta p( \phi) =  
 - 4 \ln 2 - \ln \left( 1 - \phi \right) +
\int_{1}^{\phi}
{\textrm d} \phi_1 
\left[ \frac{ \rho(\phi_1) - \phi_1 }
{\phi_1 \left( 1 - \phi_1 \right) }
\right]
\label{inter}
\end{equation}
The integrand in Eq. (\ref{inter}) is well behaved in the limit $\phi_1 \rightarrow 1$, and thus Monte Carlo Simulation and thermodynamic integration 
may be used to calculate the value of $\phi$ (and subsequently the value of $\beta \mu$) at liquid-vapor coexistence.


\begin{thebibliography}{41}

\bibitem{[11]} S. C. Glotzer and M. J. Solomon, Nat. Mat. 6, 557 (2007).
\bibitem{[12]} A. B. Pawar and I. Kretzschmar, Macromol. Rapid Comm. 31, 150 (2010).
\bibitem{[13]} E. Bianchi, R. Blaak, and C. Likos, Chem. Phys. Phys. Chem., doi 10.1039/c0cp02296a (2011).
\bibitem{[14]} J. M. Tavares, P. I. C. Teixeira, and M. M. Telo da Gama, Phys. Rev. E 80, 021506 (2009).
\bibitem{[15]} J. M. Tavares, P. I. C. Teixeira, and M. M. Telo da Gama, Mol. Phys. 107, 453 (2009).
\bibitem{[16]} Y.-S. Cho, G.-R. Yi, J.-M. Lim, S.-H. Kim, V. N. Manoharan, D. J. Pine, and S.-M. Yang, J. Am. Chem. Soc. 127, 15968 (2005).
\bibitem{[17]} Q. Chen, S. C. Bae, and S. Granick, 469, 381 (2011).
\bibitem{[18]} E. Bianchi, P. Tartaglia, E. La Nave, and F. Sciortino, J. Phys. Chem. B 111, 11765 (2007).
\bibitem{[19]} E. Bianchi, J. Largo, P. Tartaglia, E. Zaccarelli, and F. Sciortino, Phys. Rev. Lett. 97, 168301 (2006).
\bibitem{[20]} B. Ruzicka, E. Zaccarelli, L. Zulian, R. Angelini, M. Sztucki, A. Moussa, T. Narayanan, and F. Sciortino, Nat. Materials 10, 56 (2011).
\bibitem{[21]} J. Russo, J. M. Tavares, P. I. C. Teixeira, M. M. Telo da Gama, and F. Sciortino, J. Chem. Phys. 135, 034501 (2011).
\bibitem{[22]} J. Russo, J. M. Tavares, P. I. C. Teixeira, M. M. Telo da Gama, and F. Sciortino, Phys. Rev. Lett. 106, 085703 (2011).
\bibitem{[23]} J. M. Tavares, P. I. C. Teixeira, M. M. T. da Gama, and F. Sciortino, J. Chem. Phys. 132, 234502 (2010).
\bibitem{[24]} M. Wertheim, J. Stat. Phys. 35, 19, ibid. 35 (1984).
\bibitem{[25]} M. Wertheim, J. Stat. Phys. 42, 459, ibid. 477 (1986).
\bibitem{[26]} T. Tlusty and S. A. Safran, Science 290, 1328 (2000).
\bibitem{[27]} P. J. Flory, J. Am. Chem. Soc. 63, 683 (1941).
\bibitem{[28]} W. H. Stockmayer, J. Chem. Phys. 11, 45 (1943).
\bibitem{[29]} J. M. Tavares, P. I. C. Teixeira, and M. M. Telo da Gama, Phys. Rev. E 81, 010501(R) (2010).
\bibitem{[30]} F. Sciortino, E. Bianchi, J. F. Douglas and P. Tartaglia, J. Chem. Phys. 126, 194903 (2007).
\bibitem{[31]} J. M. Tavares, B. Holder and M. M. Telo da Gama, Phys. Rev E 79, 021505 (2009).
\bibitem{[32]} N. G. Almarza, J. M. Tavares and M. M. Telo da Gama, Phys. Rev. E 82, 061117 (2010).
\bibitem{[33]} N. G. Almarza, J. M. Tavares and M. M. Telo da Gama, J. Chem. Phys. 134, 071101 (2011).
\bibitem{[34]} L. G. L\'opez, D. H. Linares, and A. J. Ramirez-Pastor, Phys. Rev. E 80, 040105(R) (2009).
\bibitem{[35]} L. G. L\'opez, D. H. Linares, and A. J. Ramirez-Pastor, J. Chem. Phys. 133, 134702 (2010).
\bibitem{[36]} L. G. L\'opez, D. H. Linares, A. J. Ramirez-Pastor and S. A. Cannas, J. Chem. Phys. 133, 134706 (2010).
\bibitem{Lomba_2005} E. Lomba, C. Mart\'{\i}n, N.G. Almarza, and F. Lado, Phys. Rev E {\bf 71}, 046132 (2005).
\bibitem{Kofke} D. A. Kofke, J. Chem. Phys. 98, 4149 (1993).
\bibitem{Hoye} J. S. H{\o}ye, E. Lomba, and N.G. Almarza,  Mol. Phys. {\bf 107}, 321 (2009).
\bibitem{Almarza-Noya} N. G. Almarza and E. G. Noya, Mol. Phys. {\bf 109}, 65 (2011).
\bibitem{Almarza-Capitan} N. G. Almarza, J. A. Capit\'an, J. A. Cuesta, and E. Lomba, J. Chem. Phys. {\bf 131}, 124506 (2009).
\bibitem{Almarza-Lomba-2008} N. G. Almarza, E. Lomba, C. Mart\'{\i}n, and A. Gallardo,  J. Chem. Phys. {\bf 129}, 234504 (2008).
\bibitem{Landau_Binder} D. P. Landau and K. Binder, {\em A Guide to Monte Carlo Simulations in Statistical Physics}, 2nd ed. (Cambridge University Press), Cambridge, 2005.
\bibitem{Wilding} N. B. Wilding, Phys. Rev. E 52, 602 (1995).
\bibitem{Perez-Pellitero} J. Pérez-Pellitero, P. Ungerer, G. Orkoulas, and A. D. Mackie, J. Chem.  Phys. 125, 054515 (2006).
\bibitem{Salas} J. Salas and. A. D. Sokal,  J. Stat. Phys., {\bf 98}, 551 (2000).
\bibitem{Barma} M. Barma, and M. E. Fisher, Phys. Rev. Lett. {\bf 53}, 1935 (1984).
\bibitem{Kamienarz} G. Kamienarz and H.W.J. Bl\"ote, J. Phys. A: Math. Gen. {\bf 26}, 201 (1993).
\bibitem{Blote} H. W. J. Bl\"ote, M.P.N. den Nijs, Phys. Rev. B {\bf 37}, 1766 (1988).
\bibitem{Nienhuis} B. Nienhuis, J. Phys. A: Math. Gen. {\bf 15}, 199 (1982).
\bibitem{Bruce} A. D. Bruce, J. Phys, A: Math. Gen. {\bf 18}, L873 (1985).
\bibitem{Blote_1995} H. W. K. Bl\"ote, J. Phys. A: Math. Gen. {\bf 28}, 6289 (1995).
\bibitem{Coniglio_1977} A. Coniglio, C. R. Nappi, F. Peruggi, and L. Russo, J. Phys. A: Math. Gen. {\bf 10}, 205 (1977).
\bibitem{frenkel-smit} D. Frenkel and B. Smit, {\em Understanding Computer Simulation, From Algorithms to Applications}, 2nd ed. (Academic Press), 
New York, 2002. 

\end{thebibliography}
\end{document}